\documentclass[fleqn,usenatbib]{mnras}

\usepackage{mathptmx}

\usepackage[T1]{fontenc}

\usepackage[dvipsnames]{xcolor}

\DeclareRobustCommand{\VAN}[3]{#2}
\let\VANthebibliography\thebibliography
\def\thebibliography{\DeclareRobustCommand{\VAN}[3]{##3}\VANthebibliography}

\usepackage{graphicx}   \usepackage{amsmath}    \usepackage{amssymb}

\definecolor{ao(english)}{rgb}{0.0, 0.5, 0.0}

\colorlet{gwpurple}{blue!40!red}

\title[Relativistic outflow model of Cyg X-1 polarization]{A relativistic outflow model of the X-ray polarization in Cyg X-1}

\author[Dexter \& Begelman]{
Jason Dexter,$^{1,2}$\thanks{jason.dexter@colorado.edu} and
Mitchell C. Begelman$^{1,2}$
\\
$^{1}$JILA, University of Colorado and National Institute of Standards and Technology, 440 UCB, Boulder, CO 80309-0440, USA\\
$^{2}$Department of Astrophysical and Planetary Sciences, University of Colorado, 391 UCB, Boulder, CO 80309-0391, USA\\
}

\date{Accepted XXX. Received YYY; in original form ZZZ}

\pubyear{2023}

\begin{document}
\label{firstpage}
\pagerange{\pageref{firstpage}--\pageref{lastpage}}
\maketitle

\begin{abstract}
We propose that the polarization of the emission from the X-ray binary Cygnus X-1, measured using the {\it Imaging X-ray Polarimetry Explorer} ({\it IXPE}), is imprinted by bulk Comptonization of coronal emission in a mildly relativistic wind or jet with a hollow-cone geometry.  Models based on scattering in a static corona overlying a thin accretion disc have difficulty reproducing the relatively high polarization degree (PD $\sim 4\%$) concurrently with the low inclination ($\sim 30^\circ$) of the binary orbit.  We show that bulk outflow with a Lorentz factor $\gtrsim 1.5$ is adequate to reproduce the observed PD, with position angle parallel to the large-scale jet, provided that the scattering occurs in a conical sheath offset from the jet axis and our line of sight aligns roughly with the opening angle of the cone.  
Physically, this flow geometry could represent the entrainment of dense material near the base of an accelerating jet as it passes through the disc corona, or a slow (but still relativistic) sheath around a fast jet.  
If similar outflows are present in other X-ray binaries at higher inclination, we might expect to see still higher degrees of linear polarization $\lesssim 10\%$ with an orientation perpendicular to the jet direction.
\end{abstract}

\begin{keywords}
black hole physics -- X-rays: binaries -- accretion, accretion discs --- radiative transfer
\end{keywords}

\section{Introduction}

Accreting black holes in X-ray binary systems (BHBs) show distinct spectral and variability states \citep[e.g.,][]{remillard2006,done2007,belloni2016}. The ``soft'' states show thermally-dominated spectra peaking at a photon energy of $\simeq 1$ keV, broadly consistent with optically thick emission from a geometrically thin accretion disc \citep{shakura1973}. The spectrum in the ``hard'' states is instead an inverted power-law rising to an SED peak at $\simeq 100$ keV, thought to arise from inverse Compton scattering of lower energy seed photons in hot, optically thin plasma \citep[e.g.,][]{sunyaev1980}. An X-ray reflection contribution, including broadened iron lines, and reverberation also suggest the presence of cold material, e.g., a thin disc, even in the hard state. A significant synchrotron contribution from relativistic jet electrons may also be present \citep{markoff2001jet}.  

Despite these rich spectroscopic and variability data, the relative contributions of these components and their geometric relationships remain unclear.  For this reason, the detection of X-ray polarization from the hard state of Cygnus X-1, by the {\it Imaging X-ray Polarimetry Explorer} ({\it IXPE}) \citep{krawczynski22}, places important constraints on the geometry and mechanisms responsible for the X-ray emission in BHBs.    

The {\it IXPE} observations indicate 4\% linear polarization (LP) in the 2--8 keV band, with the position angle (PA) of the electric vector aligned with the projection of Cyg X-1's large-scale jet on the sky. The LP fraction and its PA appear to be stable despite large intensity variations in the source.  
At first glance, these observations appear to support the favored model of a scattering layer covering the inner disc and extended  perpendicular to the jet axis, which is assumed to be normal to the disc. But measurements of the binary inclination \citep{millerjones21} yield $i \approx 28^\circ$, which is too small to explain a polarization as high as 4\% for any straightforward model. If the disc is producing the observed polarization intrinsically through scattering, it would have to be oriented much closer to edge-on, and significantly misoriented with respect to both the binary orbital plane and the large-scale jet \citep{krawczynski22,zdziarski2023}. 

Here, we show that relativistic Comptonization in a moderately relativistic outflow could produce a larger polarization degree with the PA found in Cyg X-1. \citet[][BS87]{begelman87} showed that scattered radiation from a jet containing {\it cold} electrons --- i.e., electrons with random Lorentz factors smaller than the bulk jet Lorentz factor, $\Gamma$ --- can be highly polarized even when the background radiation field is isotropic and unpolarized. The reason is that the relativistically moving electrons see highly anisotropic radiation beamed from their direction of motion. An observer viewing the jet from an angle $\sim \Gamma^{-1}$ with respect to the direction of motion sees photons that are scattered by about $90^\circ$ in the electron frame, which can be nearly 100\% polarized. For jets with scattering concentrated along the axis, BS87 found that the observed polarization could be 10's of percent but  that the PA would be perpendicular to the projected jet axis on the sky. However, a jet with opening angle $\theta_j$ and scattering concentrated along its walls --- a {\it hollow-cone} jet --- could produce significant parallel polarization when viewed from  an angle $\theta_j \pm \Gamma^{-1}$. 

We propose that the high polarization observed in Cyg X-1 is imprinted by scattering off a moderately relativistic outflow as it passes through the corona near its base. The onset of  relativistic effects is remarkably insensitive to bulk Lorentz factor, and we find that $\Gamma \gtrsim 1.5$ is sufficient to produce the observed polarization. At low Lorentz factor, the cold outflow can imprint a large degree of linear polarization without significantly modifying the spectrum of incident radiation.

We describe an analytic model of relativistic Comptonization (Section 2) and show that a collimated, mildly relativistic outflow provides a natural explanation for the observed X-ray polarization measured from Cyg X-1 (Section 3) without requiring a misalignment between the accretion flow and binary orbit. The outflow is mass-loaded and produces $10-20\%$ of the observed X-ray flux, but carries relatively little mass and energy. In Section 4, we discuss the implications of the model for the physical origin of the corona, and make predictions for future X-ray polarization observations from similar sysems.

\begin{figure}
   \includegraphics[width=\columnwidth]{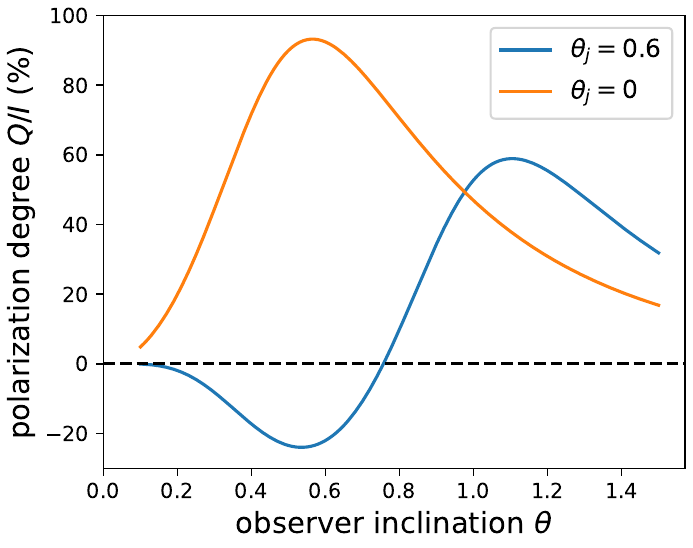}
    \caption{Linear polarization $Q/I$ as a function of observer's inclination for relativistic Comptonization ($\Gamma = 2$) from a delta-function outflow oriented along the polar axis ($\theta_j = 0$) and inclined from the axis ($\theta_j = 0.6$). The maximal polarization degree is reached for the polar case, with an orientation perpendicular to the jet ($Q > 0$). For a hollow-cone outflow viewed from the inside ($\theta < \theta_j$), the polarization can instead be parallel to the jet ($Q < 0$) with a degree of $\approx 20\%$.}
    \label{fig:sample}
\end{figure}

\begin{figure}
   \includegraphics[width=\columnwidth]{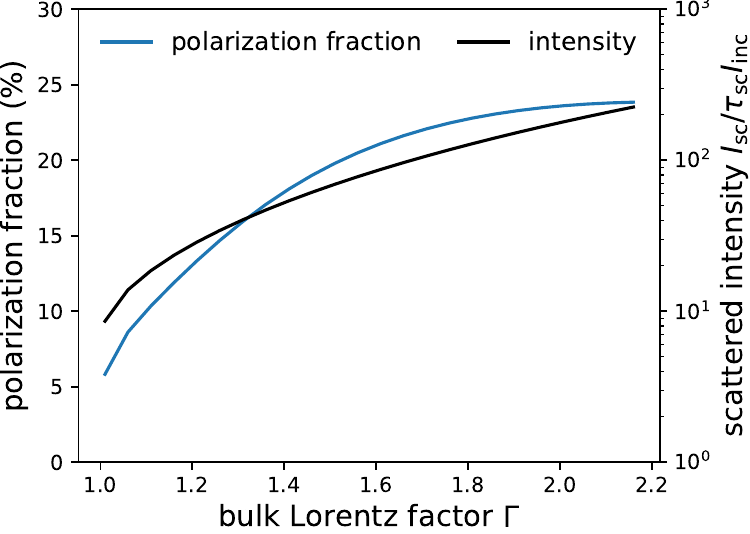}
    \caption{Linear polarization fraction and scattered intensity from relativistic Comptonization as a function of the outflow Lorentz factor for the delta-function hollow-cone case with an isotropic seed photon distribution and $\theta_j = 0.6$, $\theta = 0.5$. The polarization fraction rapidly increases with outflow velocity due to the relativistic beaming effect, exceeding $20\%$ for $\gamma \gtrsim 1.5$ or $\beta \gtrsim 0.75$.}
    \label{fig:gamma}
\end{figure}

\begin{figure*}
   \begin{tabular}{cc}
   \includegraphics[width=\columnwidth]{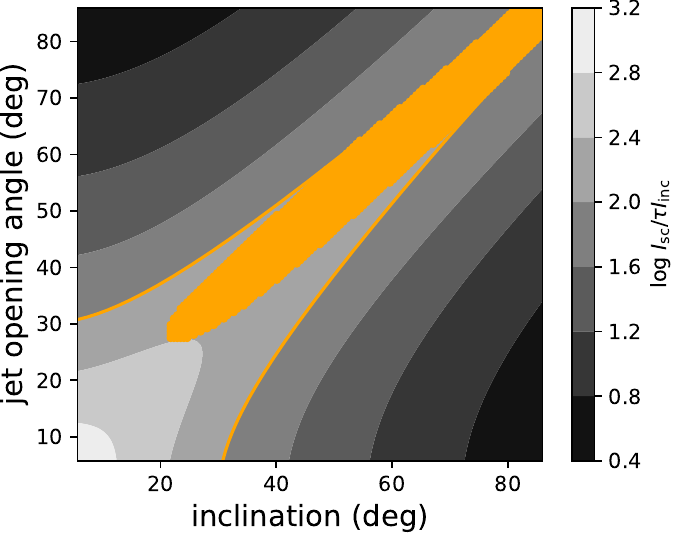} &
   \includegraphics[width=\columnwidth]{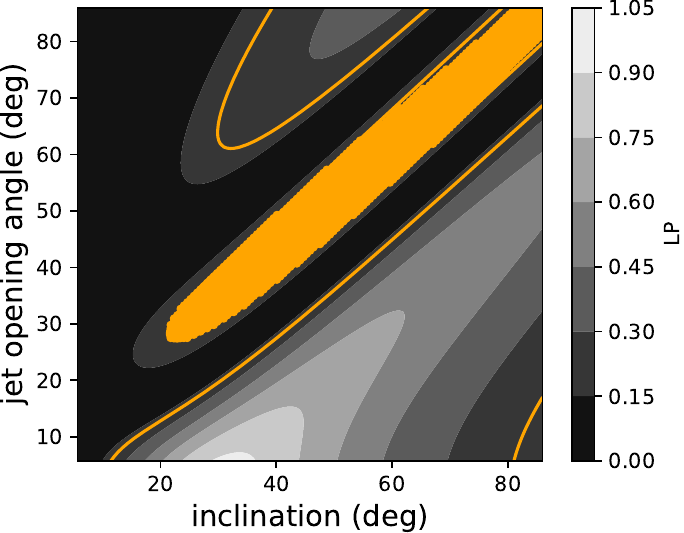}
   \end{tabular}
    \caption{Comptonized intensity fraction (left) and linear polarization fraction (right) for a relativistic outflow as a function of opening angle $\theta_j$ and inclination angle $\theta$, for a delta-function hollow cone jet with $\Gamma = 2$. The thin orange lines show contours corresponding to the individual constraints $I_{\rm sc}/\tau I_{\rm inc} > 100$ (left) and LP $> 0.2$ (right), while the orange shaded region shows models satisfying all constraints including that the polarization angle lie along the jet direction. Viable solutions exist for a large range of $\theta_j \sim \theta$, including $\theta \simeq 30^\circ$ corresponding to the binary inclination angle of Cyg X-1.}
    \label{fig:xraypol}
\end{figure*}

\section{A relativistic bulk Comptonization model for Cyg X-1}
\label{sec:relcompton}

BS87 developed an analytic model of polarized radiation from inverse Compton scattering in a jet geometry. In particular, they assumed cold electrons such that $k T_e/m_e c^2 \ll \Gamma$, with $\Gamma$ the bulk Lorentz factor of the outflow. We further consider the case of an isotropic incident radiation field, e.g., that the scattering region is isotropically irradiated, and that the seed photons are distributed as a power-law in energy, $I_{\rm inc} = C \nu^{-s}$. For simplicity we fix $s=1$ throughout. In the rest frame of the outflow, incident intensity is
 \begin{equation}
     I'_{\rm inc} = C (\nu')^{-s} \Gamma^{-3-s} (1+\beta \cos \rho')^{-3-s},
 \end{equation}
where $\beta = v/c$ is the outflow velocity and $\Gamma$ its Lorentz factor, $s$ is the spectral index, $\nu'$ is the frequency in the comoving frame and and $\rho'$ is the polar angle between the line of sight and outflow as viewed in the outflow's rest frame.
 
 In this case, the emergent Stokes parameters can be written as (BS87 equation 15),
\begin{eqnarray}\label{fulleqn}
\frac{I_{\rm sc}}{\tau_{\rm sc} I_{\rm inc}}  &=& \frac{3}{16\pi} \int_1^{\infty} {d\Gamma \over \Gamma^{6+2s} } \int d\Omega(\theta_e,\phi_e) \frac{n(\Gamma,\hat{e})/n}{\left(1-\beta \cos \rho_0\right)^{2+s}} \nonumber\\ & & \int d\Omega (\rho', \alpha') \frac{1+\cos^2 w'}{(1+\beta \cos \rho')^{3+s}},  \\
\frac{Q_{\rm sc}}{\tau_{\rm sc} I_{\rm inc}} &=& \frac{3}{16\pi} \int_1^{\infty} {d\Gamma \over \Gamma^{6+2s} } \int d\Omega(\theta_e,\phi_e) \frac{n(\Gamma,\hat{e})/n}{\left(1-\beta \cos \rho_0\right)^{2+s}} \nonumber \\ & & \cos 2\xi  \int d\Omega (\rho', \alpha') \frac{1-\cos^2 w'}{(1+\beta \cos \rho')^{3+s}}, 
\end{eqnarray}

\noindent where we have defined a series of angles describing the scattering geometry. The angle $w'$ is the scattering angle of the photon, $\rho$ and $\rho_0$ are the angles between the electron velocity and the incident and scattered photon, $\xi$ is the angle between the electron velocity and the polar axis as projected on the sky, and $U_{\rm sc} = 0$ for the axisymmetric electron distributions considered here. Primed quantities are measured in the electron rest frame. In this convention, the polarization angle is given by EVPA $= 1/2 (\pi - \tan^{-1} U/Q)$. In particular, $Q > 0$ corresponds to polarization lying in the equatorial plane while $Q < 0$ corresponds to polarization aligned with the polar axis.

We consider two cases of electrons concentrated around a polar angle $\theta_j$: a purely beamed distribution $n(\Gamma,\theta_e)/n = \delta(\theta_e-\theta_j)$ and one with a Gaussian distribution of angular width $\Delta \theta$, $n(\Gamma,\theta_e)/n \propto e^{-(\theta_e-\theta_j)^2/2\Delta \theta^2}$.

\autoref{fulleqn} is accurate for all $\Gamma$. In the relativistic limit $\Gamma \gg 1$, BS87 developed a ``head-on'' approximation, noting that the scattering is dominated by photons approaching the electrons head-on in the electron rest frame. In that case, $\rho' \approx \pi$ and the above equations simplify to BS87 equation 23:

\begin{eqnarray}\label{headon}
\frac{I_{\rm sc, ho}}{\tau_{\rm sc} I_{\rm inc}} &=& \frac{3}{2} \frac{2^s}{2+s} \int_1^{\infty} {d\Gamma \over \Gamma^{2}} \int d\Omega(\theta_e,\phi_e) \frac{n(\Gamma,\hat{e})}{n}\frac{1+\cos^2 \rho'_0}{\left(1-\beta \cos \rho'_0\right)^{2+s}},  \nonumber \\   \\
\frac{Q_{\rm sc, ho}}{\tau_{\rm sc} I_{\rm inc}} &=& \frac{3}{2} \frac{2^s}{2+s} \int_1^{\infty} {d\Gamma \over \Gamma^{2} } \int d\Omega(\theta_e,\phi_e) \frac{n(\Gamma,\hat{e})}{n}\frac{(1-\cos^2 \rho'_0) \cos 2 \xi}{\left(1-\beta \cos \rho'_0\right)^{2+s}}. \nonumber \\
\end{eqnarray}

\autoref{fig:sample} shows the resulting linear polarization for cases of polar ($\theta_j = 0$) and ``hollow-cone'' ($\theta_j = 0.6$) electron distributions $n_{\hat e} \propto \delta(\theta_e-\theta_j)$ with a bulk Lorentz factor $\gamma = 2$. The result shown is for a full calculation -- in both cases, the linear polarization in the ``head-on'' approximation is accurate to within 20\%. As shown in BS87, the linear polarization fraction nears unity for the case of polar electrons, with the polarization oriented perpendicular to the axis. For a hollow-cone electron distribution, the polarization angle flips from oriented parallel to the axis ($Q_{\rm sc} < 0$) when viewed inside the cone to perpendicular ($Q_{\rm sc} > 0$) when viewed outside. The maximum polarization degree is $\approx 25\%$ for polarization aligned with the axis. 

\autoref{fig:gamma} shows the polarization fraction and relative scattered intensity as a function of the outflow Lorentz factor for a full calculation where an  observer at $\theta = 0.5$ looks down at a narrow (delta-function) hollow-cone outflow with $\theta_j = 0.6$. Relativistic effects sharply increase both the scattered intensity and linear polarization fraction even at modest Lorentz factors. For $\Gamma \gtrsim 1.5$, the polarization fraction approaches its maximum value of $\gtrsim 25\%$.

We calculate these quantities in the head-on approximation over a wide range of outflow angle $\theta_j$ and observer inclination $\theta$. For {\it IXPE} observations of Cyg X-1, we apply constraints of LP $\ge 20\%$ (for the measured 4\% polarization fraction in Cyg X-1, which requires that at least 20\% of the observed flux be scattered by the outflow), a polarization angle lying along the jet, and a relative scattered intensity $I_{\rm sc} / \tau I_{\rm inc} \ge 100$. The observed scattered flux fraction is then $F_{\rm sc} / F_{\rm inc} = \tau f_{\Omega_j} (I_{\rm sc} / \tau I_{\rm inc})$, where $f_{\Omega_j}$ is the fractional solid angle subtended by the outflow. For a purely radial outflow, $f_{\Omega_j} \simeq 1/2 [\cos{(\theta_j-\Delta \theta)} - \cos{(\theta_j+\Delta \theta})]$. For a collimated outflow we expect $f_{\Omega_j} \sim 0.1-0.2$. 

Results are shown in \autoref{fig:xraypol} for a bulk Lorentz factor $\Gamma = 2$. The major constraint for reproducing the observed X-ray polarization in Cyg X-1 is that the observer should be looking along the hollow cone of the jet, $\theta \sim \theta_j$. Given that constraint, a wide range of inclination and opening angles can match the data. In particular, solutions are available for the observed binary system inclination angle of $\simeq 30^\circ$ provided that we are viewing within $\simeq 10^\circ$ of the outflow direction.

\section{Discussion}
\label{sec:discussion}

{\it IXPE} observations of linear polarization in Cyg X-1 show a relatively high polarization fraction of $4\%$ \citep{krawczynski22}. We have shown that bulk Comptonization in a mildly relativistic, collimated outflow can produce large linear polarization fractions consistent with the observations if $\approx 20\%$ of the observed X-ray luminosity is produced by scattering. For a polarization angle along the jet axis, the outflow should be a ``hollow cone'' observed from inside, e.g., $\theta \lesssim \theta_j$. Unlike non-relativistic Comptonization models, relativistic bulk Comptonization can explain the {\it IXPE} results without requiring that the accretion flow be misaligned with the binary angular momentum.

We have explored the parameter space of bulk Comptonization in an analytic framework following BS87. The major assumption of this treatment is that the seed photon distribution is isotropic.
This basic picture would still apply for a mildly anisotropic radiation field, provided that aberration due to the bulk Lorentz factor dominates the angular intensity distribution. The linear polarization fraction increases rapidly with Lorentz factor, such that only moderate outflow velocities are required ($\Gamma \gtrsim 1.5$). Given size constraints on the X-ray corona, this implies that the outflow is launched from small radius $r \lesssim 10 r_g$. The model otherwise does not favor any particular model for the geometry of the X-ray corona. The outflow velocity is similar to the escape speed from close to the black hole. The hollow-cone geometry would seem natural if the outflow borders a relativistic jet along the polar axis, which is not mass loaded and so does not produce significant amounts of Comptonized radiation. Such an outflowing ``jet wall'' region is commonly found in GRMHD simulations \citep[e.g.,][]{hirose2004,mckinney2006}, and may correspond to the limb-brightened region seen in ultra-high resolution images of radio jets of M87 and Cen A \citep{hada2011,ehtcena}.

We have also assumed purely bulk Comptonization due to ``cold'' electrons, e.g., $\Gamma \gg \Theta_e$ with $\Theta_e = k T_e / m_e c^2$. This is appropriate for $\Gamma \gtrsim 1.5$, even if the coronal temperature is as high as $T_e \sim 10^9$ K. The walls of the cone should also be relatively narrow. For our fiducial parameters, a velocity width of $\Delta \theta = 0.2$ decreases the polarization degree from $20\%$ to $10\%$. A relatively narrow-angle outflow could naturally cover a sufficiently large fractional solid angle $f_\Omega \sim 0.1$ to explain the total observed polarized flux.

The incident radiation field in our model is assumed to be isotropic, e.g., the outflow is embedded in the corona. \citet{poutanen23} showed that similarly high polarization degrees can be obtained from a mildly relativistic outflow irradiated by an anisotropic incident radiation field. In our scenario, this would reduce the overall reflected flux since fewer photons would be scattered towards the observer. However, the polarization degree might also be higher.

In a fiducial case with $\theta \sim \theta_j$ and $\Gamma = 1.5$, we require $\tau_{\rm sc} \simeq 0.1$ in order to produce $\gtrsim 20\%$ of the total X-ray flux. Due to its low density, the outflow mass flux $\dot{M} \lesssim 10^{-11} f_\Omega \, M_\odot \, \rm yr^{-1}$ is still a small fraction $\sim 10^{-3}$ of the mass accretion rate. In Cyg X-1, a non-thermal component produces a hard power law tail with $\approx 10\%$ of the total flux. This is similar to what we infer from relativistic Comptonization. It is possible that the outflow considered here could contain a non-thermal particle distribution extending to higher energies. Alternatively, that radiation could also come from a jet extending out to large radius \citep[e.g.,][]{russell2014}.

\emph{IXPE} has recently detected a high degree of linear polarization in the soft state of the X-ray binary 4U 1630-47, which is difficult to reconcile with standard disk atmosphere models \citep{ratheesh2023,kushwaha2023}. A collimated, relativistic disk wind may also be present in such states, and Comptonization of the disk emission in the outflow as modeled here could produce the observed linear polarization while still retaining a thermal spectrum.

Our results would imply that the corona consists of a thermal population producing most of the X-ray flux, along with a moderately relativistic outflow (disk-wind or jet wall). Since all of the polarized flux is produced in a collimated, relativistic outflow our model does not favor any particular coronal geometry. If other systems have similar outflows, their observed polarization with {\it IXPE} should depend strongly on viewing geometry. For systems viewed at low inclination, we should be looking inside the outflow. Then the maximum linear polarization from Comptonization is $\approx 25\% f_{\rm outflow}$ with a polarization angle aligned with the rotation axis. For systems observed at higher inclination, the polarization angle should instead be perpendicular to the rotation axis, potentially with much higher degree of linear polarization $\approx 90\% f_{\rm outflow}$. Measuring both the system orientation and polarization degree thus provides a test for the presence of relativistic outflows from the near horizon region in black hole X-ray binary systems.

\section*{Acknowledgements}
This work was supported in part by NASA Astrophysics Theory Program grants NNX16AI40G, NNX17AK55G, and 80NSSC20K0527, Fermi award 80NSSC21K2027, Chandra award  TM3-24003X, and by an Alfred P. Sloan Research Fellowship (JD). 

\section*{Data Availability}

The Mathematica notebook used to carry out the Comptonization calculations presented here will be made available upon request.

\bibliographystyle{mnras}

\bsp    \label{lastpage}
\end{document}